\documentclass[aps,prd,onecolumn,groupedaddress,showpacs,nofootinbib,amssymb]{revtex4}
\usepackage[dvips]{graphicx}
\usepackage{amssymb}
\usepackage{amsmath}
\usepackage{graphicx}
\usepackage{amsfonts}
\usepackage{bm}
\usepackage{latexsym,float}
\usepackage{epsfig}
\usepackage{epstopdf}
\usepackage{amssymb}
\usepackage{verbatim}
\usepackage{multirow}
\usepackage{color}

\begin{document}

\title{Correspondence of $F(R)$ Gravity Singularities in Jordan and Einstein Frames}

\author{Sebastian Bahamonde}
\email{sebastian.beltran.14@ucl.ac.uk}
\affiliation{Department of Mathematics, University College London,
    Gower Street, London, WC1E 6BT, UK}

\author{S.~D.~Odintsov}
\email{odintsov@ieec.uab.es}
\affiliation{Institut de Ciencies de lEspai (IEEC-CSIC),
    Campus UAB, Carrer de Can Magrans, s/n, 08193 Cerdanyola del Valles, Barcelona, Spain}
\affiliation{ICREA, Passeig LluAs Companys, 23,
    08010 Barcelona, Spain}
 
\author{V. K. Oikonomou}
\email{v.k.oikonomou1979@gmail.com,voiko@sch.gr}
\affiliation{Tomsk State Pedagogical University, 634061 Tomsk}
\affiliation{Laboratory of Theoretical Cosmology, Tomsk State University of Control Systems\\ 
and Radioelectronics (TUSUR), 634050 Tomsk, Russia }    
    
\author{Matthew Wright}
\email{matthew.wright.13@ucl.ac.uk}
\affiliation{Department of Mathematics, University College London,
    Gower Street, London, WC1E 6BT, UK  }
    
%\date{\today}

\begin{abstract}
We study the finite time singularity correspondence between the Jordan and Einstein frames for various $F(R)$ gravity theories. Particularly we investigate the ordinary pure $F(R)$ gravity case and the unimodular $F(R)$ gravity cases, in the absence of any matter fluids. In the ordinary $F(R)$ gravity cases, by using specific illustrative examples, we show that it is possible to have various correspondences of finite time singularities, and in some cases it is possible a singular cosmology in one frame might be non-singular in the other frame. In the unimodular $F(R)$ gravity case, the unimodular constraint is affected from the conformal transformation, so this has an effect on the metric we choose. Moreover, we study the Einstein frame counterpart theory of the unimodular $F(R)$ gravity case, and we investigate the correspondences of the singularities in the two theories by considering specific illustrative examples. Finally, a brief dynamical system analysis is performed for the vacuum unimodular $F(R)$ gravity and we demonstrate how the dynamical system behaves near the future Big Rip singularity.
\end{abstract}


\maketitle

\section{Introduction}

Cosmological singularities are timelike singularities in contrast to spacelike singularities, which occur in the case of compact astrophysical objects, like black holes \cite{Virbhadra:2002ju}. The finite time cosmological singularities affect the whole three dimensional spacelike hypersurface defined by the time instance that these occur, and these were concretely classified for the first time in \cite{Nojiri:2005sx}. In the classification scheme of Ref. \cite{Nojiri:2005sx}, there are four different singularity types, among which the Big Rip \cite{ref5} is the most severe from a phenomenological point of view, since all the physical quantities which are defined on the spacelike hypersurface defined by the singularity, severely diverge. On the antipode of the Big Rip singularity lies the so-called Type IV singularity \cite{Nojiri:2005sx,noo1,noo2,noo3,noo6}, which is a soft singularity, since all the physical quantities are finite at the time instance that the singularity occurs, but the higher derivatives of the Hubble rate diverge. From a first glance, the Type IV singularity seems not to affect the cosmological system, since the singularity does not alter the physics of the three dimensional hypersurfaces, however the occurrence of the Type IV singularity, crucially affects the dynamical evolution of the cosmological system, as was explicitly demonstrated in \cite{noo2,noo3}, and in some cases this might have interesting consequences. Particularly, it is possible that the existence of a Type IV singularity triggers a graceful exit from inflation, due to a dynamical instability caused by the Type IV singularity, see \cite{noo3} for details. In addition, in some cases, a Type IV singularity can be constrained by physical processes, for example in Ref. \cite{noo6} it was shown that the gravitational baryogenesis can constrain the form of the Type IV singularity.

In the literature there exists a class of singularities, known as sudden singularities, which were firstly studied in Ref. \cite{barrownew}, and later developed in Refs. \cite{Barrow:2004xh,barrow}. The Type IV singularity belongs to this class of singularities, and the possibility of having singular inflation with sudden singularities was studied in Ref. \cite{Barrow:2015ora}.  A particular interesting feature of milder singularities is that geodesics incompleteness does not occur for milder singularities, in contrast with the Big Rip case, where geodesics incompleteness occurs, hence in the case of mild singularities, a smooth passage of the Universe through them is guaranteed.       

The finite time cosmological singularities can be consistently described by $F(R)$ modified gravity models \cite{reviews1}, since the theoretical framework of modified gravity offers many possibilities for realization of cosmological scenarios which are exotic for standard Einstein-Hilbert gravity. The purpose of this paper is to investigate in detail the correspondence of the finite time singularities between the Einstein and Jordan frames, in the context of various $F(R)$ gravity frameworks. For a recent work using scalar-tensor theories, see \cite{vernov}. It is a well known fact that every Jordan frame $F(R)$ gravity has an Einstein frame counterpart, related to each other by a conformal transformation \cite{reviews1}, so the basic issue we will address with this paper is how the finite time cosmological singularities are transformed from one frame to the other. As we will demonstrate, it is possible that the mild singularities in one frame correspond to crushing type singularities in the other frame, under the conformal transformation. Of course, it is expected that the $F(R)$ gravity framework plays some crucial role in the study, so we shall investigate the Jordan-Einstein frame correspondence in various $F(R)$ gravity variants. Specifically, we shall use the recently developed unimodular $F(R)$ gravity \cite{unimodular1,unimodular4} and also the usual vacuum $F(R)$ framework. For simplicity we focus on the case that the $F(R)$ gravity is described by a power law function of the Ricci scalar, of the form $R^{-n}$, but in principle the study can be extended in other cases too. After providing some essential information about finite time singularities and the essential features of the conformal transformation relating the Jordan and Einstein frames of $F(R)$ gravity, we investigate how certain singularities behave under the conformal transformation. We also discuss how specific cosmological evolutions are transformed under this transformation. In addition, we discuss the same issues in the context of unimodular $F(R)$ framework. By using several illustrative examples, we demonstrate that there exist several patterns of transformations from frame to frame, depending also to the form of the $F(R)$ gravity. Also, in the case of the unimodular $F(R)$ gravity, we develop the necessary formalism that relates the Jordan and Einstein frames. Note that it is known that even in the case of no singularities, there is no physical
equivalence of Jordan and Einstein frames, see for example \cite{newref,Briscese:2006xu}.

This paper is organized as follows: In section \ref{jordanframefr}, we present some fundamental information about finite time singularities and we study in detail the $F(R)$ gravity case. After demonstrating in brief how to obtain the Einstein frame counterpart, we investigate how the cosmological singularities behave under the conformal transformation, for the case that the $F(R)$ gravity is of the power law type $R^{-n}$.  In section \ref{unimodularfr}, we briefly present the unimodular $F(R)$ gravity framework and also the corresponding Einstein frame counterpart of this theory. We also give an account on the unimodular theory of a canonical scalar field, and we use some illustrative examples in order to demonstrate explicitly how the singularities are transformed from frame to frame. In section \ref{phasestructurese} we present in brief the phase structure of the dynamical system corresponding to the vacuum unimodular $F(R)$ gravity and we discuss certain features of the dynamical system near the Big Rip singularity. Finally, the concluding remarks follow in the end of the paper. 

The geometric background we shall assume in this paper is described by a flat Friedmann-Robertson-Walker (FRW) metric, with line element,
\begin{equation}
\label{metricfrw} ds^2 = - dt^2 + a(t)^2 \sum_{i=1,2,3}
\left(dx^i\right)^2\, ,
\end{equation}
with $a(t)$ denoting as usual the scale factor. In addition, the connection is assumed to be the Levi-Civita connection, which is a torsion-less, symmetric, and metric compatible affine connection. Physical quantities in the Einstein frame will be denoted with a tilde $\tilde{}$, whereas quantities in the Jordan frame will not.

\section{$F(R)$ gravity}\label{jordanframefr}

Before we get into the core of our study, it is worth recalling at this point some essential information with regards to the finite time cosmological singularities. The detailed classification of these was firstly done in Ref. \cite{Nojiri:2005sx}, and according to which classification, there are four types of finite time singularities. The classification criteria of Ref. \cite{Nojiri:2005sx}, were the scale factor, the energy density, the pressure density and the higher derivatives of the Hubble rate. Depending on the behavior of these physical quantities on the spacelike three dimensional hypersurface corresponding to the time instance that the singularity occurs, the classification goes as follows,
\begin{itemize}
\item Type I (``the Big Rip Singularity'') : This is the most severe type of finite time cosmological singularities, being a singularity of the crushing type. This happens when, as the cosmic time approaches a certain time $t \to t_s$, the effective energy density, the scale factor and the effective pressure all diverge, so that  $a \to \infty$, $\rho_\mathrm{eff} \to \infty$, and $\left|p_\mathrm{eff}\right| \to \infty$ when $t \to t_s$. For more details about the Big Rip Singularity, we refer to the following set of important papers.~\cite{Nojiri:2005sx,ref5}
\item Type II (the ``Sudden Singularity'') \cite{barrow}: This type of singularity appears when, as the cosmic time  $t \to t_s$, the scale factor $a$ and the effective energy density $\rho_{\mathrm{eff}}$ remain bounded, that is $a \to a_s$, $\rho_{\mathrm{eff}} \to \rho_s$, where both $a_s,\rho_s<\infty$, but the effective pressure $p_\mathrm{eff}$ diverges. This means $\left|p_\mathrm{eff}\right| \to \infty$ as $t \to t_s$. This can occur when the second derivatives of the scale factor diverge.
\item Type III: This type of singularity is more severe than the Type II singularity, and appears when, as the cosmic time  $t \to t_s$, the scale factor $a$ remains bounded, that is $a \to a_s$, but this time both the effective energy density and the effective pressure diverge, meaning $\rho_\mathrm{eff} \to \infty$ and $\left|p_\mathrm{eff}\right| \to \infty$. This can happen when the first and second derivatives of the scale factor diverge.
\item Type IV : This is the least severe of the types of cosmological singularities studied here. This type of singularity appears when, as the cosmic time $t \to t_s$, the scale factor, the effective energy density and the corresponding effective pressure all remain bounded, $a \to a_s$, $\rho_\mathrm{eff} \to \rho_s$ and $\left|p_\mathrm{eff}\right| \to p_s$ , but the second or higher derivatives of the Hubble rate diverge. For more on this type of singularity, we refer the reader to the following~\cite{Nojiri:2005sx,noo1,noo2,noo3,noo6}.
\end{itemize}
Having this classification scheme in mind, let us proceed to description of the $F(R)$ gravity in the Jordan frame and we investigate how these singularities are transformed to each other with a conformal transformation.

The vacuum $F(R)$ gravity action in the Jordan frame is,
\begin{align}\label{jfr}
S = \frac{1}{2\kappa^2}\int d^4 x \, \sqrt{-g}F(R) \ .
\end{align}
If we vary this action with respect to the metric and assume a flat FRW metric, we find the following field equations in the absence of matter
\begin{eqnarray}
\label{JGRG15}
0 &=& -\frac{F(R)}{2} + 3\left(H^2 + \dot H\right) F'(R)
- 18 \left( 4H^2 \dot H + H \ddot H\right) F''(R)\,,\\
\label{Cr4b}
0 &=& \frac{F(R)}{2} - \left(\dot H + 3H^2\right)F'(R)
+ 6 \left( 8H^2 \dot H + 4 {\dot H}^2 + 6 H \ddot H + \dddot H\right) F''(R)
+ 36\left( 4H\dot H + \ddot H\right)^2 F'''(R)\,,
\end{eqnarray}
where primes and dots denote derivation with respect to the Ricci scalar $R$ and the cosmic time $t$ respectively.

We now conformally transform in order to obtain the scalar-tensor Einstein frame counterpart theory. In order to do this, we introduce the auxiliary fields $A$ and $B$, so that we can write the equivalent action of the action (\ref{jfr}), which reads,
\begin{align}
S= \frac{ 1}{2\kappa^2}  \int  d^4 x \, \sqrt{-g} \left\{B(R-A)+F(A) \right\} \,. \label{F(R)aux}
\end{align}
Varying this action with respect to the auxiliary scalar $B$, results in the condition $A=R$, and hence we recover the action~(\ref{jfr}). We can eliminate the auxiliary field $B$ from this action by varying with respect to $A$, and in effect we end up to the condition $B=F'(A)$, so the action takes the equivalent form,
\begin{align}
S= \frac{ 1}{2\kappa^2}  \int  d^4 x \, \sqrt{-g} \left\{F'(A)(R-A)+F(A) \right\} \,  \label{F(R)aux2}\,.
\end{align}
By conformally transforming the metric, we can obtain a minimally coupled scalar-tensor theory, which is called the Einstein frame scalar-tensor theory. We use a particular conformal factor of the form,
\begin{align}
\hat{g}_{\mu\nu}=\frac{1}{ F'(A)} g_{\mu\nu},
\end{align}
which modifies the Ricci scalar as $R\rightarrow \hat{R}$. After performing this conformal transformation, and upon defining a new scalar field $\sigma$ in terms of the auxiliary scalar field $A$,
\begin{align}
\sigma=-\ln F'(A), \label{Asigma}
\end{align}
the action~(\ref{F(R)aux2}) takes the following form,
\begin{align}\label{scteact}
S = \frac{1}{2\kappa^2}\int d^4 x \, \sqrt{-\hat{g}} \left\{\hat{R} -
\frac{3}{2}\hat{g}^{\mu\nu} \partial_{\mu} \varphi \partial_{\nu} \varphi  - V(\varphi)\right\} \, , 
\end{align}
with the potential $V(\sigma)$ being equal to,
\begin{align}\label{pot}
V(\sigma)=\frac{A}{F'(A)}-\frac{F(A)}{F'(A)^2}\,.
\end{align}
Note that the potential (\ref{pot}) can be easily rewritten in terms of the scalar field $\sigma$, by inverting the relation~(\ref{Asigma}). Hence the resulting Einstein frame scalar-tensor theory corresponding to the Jordan frame $F(R)$ gravity (\ref{jfr}) is given by the action of Eq. (\ref{scteact}). Finally we can rescale the scalar field $\sigma$ to put it into the canonical form by introducing
\begin{align}
\varphi= \sqrt{\frac{3}{2\kappa^2}}\sigma.
\end{align}

Conversely, by starting with the following scalar-tensor canonical scalar field action, 
\begin{align}\label{scalcan}
S = \int d^4 x \, \sqrt{-\hat{g}} \left\{\frac{\hat{R}}{2\kappa^2} -
\frac{1}{2} \hat{\partial_{\mu}} \varphi \hat{\partial}^{\mu} \varphi  - V(\varphi)\right\} \, , 
\end{align}
we can transform it to a Jordan frame $F(R)$ gravity theory. By assuming a flat FRW metric, the Friedman equations corresponding to the action (\ref{scalcan}) are equal to,
\begin{align}
3\tilde{H}^2=\frac{1}{2}\dot{\phi}^2+V\, , \\
3\tilde{H}^2+2\dot{\tilde{H}}=-\frac{1}{2}\dot{\phi}^2+V\, .
\end{align}
Now let us map to a modified gravity $F(R)$ theory. This means we must perform the conformal transformation $g_{\mu\nu}\rightarrow e^{\pm\sqrt{\frac{2}{3}}\kappa\phi}\hat{g}_{\mu\nu}$, so that the FRW metric (\ref{metricfrw}) becomes,
\begin{align}
ds_{F(R)}^2=e^{\pm\sqrt{\frac{2}{3}}\kappa\phi}\left(-d\tilde{t}^2+ \tilde{a}(\tilde{t})^2\sum_{i=1,2,3} \left(dx^i\right)^2 \right) ,
\end{align} 
and in order to obtain a FRW metric in the $F(R)$ gravity frame, we introduce a new time coordinate $\tilde{t}$ variable, given by solving the following equation,
\begin{align}
dt=e^{\pm\frac{1}{2}\sqrt{\frac{2}{3}}\kappa\phi} d\tilde{t}\,,
\end{align}
the solution of which, $t=f(\tilde{t})$, is an increasing function. Certain problems occur if the function $f(\tilde{t})$ contains singularities. The range of the values of the cosmic time $\tilde{t}$ might get mapped to a different range in the $t$ coordinate. Assume that the range of the values of the cosmic time $\tilde{t}$ is the interval $[\tilde{t}_1,\tilde{t}_2]$, with the scale factor at $\tilde{t}=\tilde{t}_1$ being equal to, $\tilde{a}(\tilde{t}_1)=0$ and at $\tilde{t}=\tilde{t}_2$, the scale factor being equal to, $\tilde{a}(\tilde{t}_2)=0$ or $\tilde{a}(\tilde{t}_2)=\infty$. The latter situation at $\tilde{t}=\tilde{t}_2$, corresponds to a Big Crunch and to a  Big Rip respectively, in which case with potentially $\tilde{t}_1=-\infty$ and/or $\tilde{t}_2=\infty$, so that these singularities effectively do not occur, a fact which crucially depends on the particular form of the scale factor at hand. The new range of $t$ variable will be $[f(\tilde{t}_1),f(\tilde{t}_2)]$, assuming $\phi(\tilde{t})$ is regular everywhere over the range $[\tilde{t}_1,\tilde{t}_2]$. The scale factor in terms of the $t$-variable is determined by the following equation,
\begin{align}
a(t(\tilde{t}))=e^{\pm\frac{1}{2}\sqrt{\frac{2}{3}}\kappa\phi}\tilde{a}(\tilde{t})\, ,
\end{align}
which upon differentiation with respect to $\tilde{t}$, it yields,
\begin{align}
\frac{da(t(\tilde{t}))}{dt}=\pm\sqrt{\frac{2}{3}}\frac{\kappa\dot{\phi}}{2}\tilde{a}(\tilde{t})+\dot{\tilde{a}}(\tilde{t})\,.
\end{align}
Therefore, in the Jordan frame, the scale factor $a$ is an increasing function if and only if the following condition holds true,
\begin{align}
\tilde{H}>\mp\sqrt{\frac{2}{3}}\frac{\kappa\dot{\phi}}{2}
\end{align}
in the original frame, and decreasing if the opposite inequality holds true.

\subsection{Power Law Cosmology}

In order to demonstrate the correspondence of finite time singularities between the Einstein and Jordan frames, we shall use some simple but illustrative examples. In the following we will work in units such that the gravitational coupling constant $\kappa=1$. We start off with the power law cosmology, described by the following scale factor,
\begin{align}
\tilde{a}(\tilde{t})=\tilde{a}_0 (\tilde{t}/\tilde{t}_0)^p \, ,
\end{align}
with $\tilde{t}_0$ being some fiducial time and $p$ a positive real free parameter. Such a power law scale factor is a solution to the Friedmann equations in the Einstein frame scalar-tensor theory when the potential is of the exponential form. In such a case the scalar field behaves as,
\begin{align}
\phi=\pm \sqrt{2p} \ln (\tilde{t}/\tilde{t}_0)\,.
\end{align}
In this case $\tilde{t}_1=0$ and $\tilde{t}_2=\infty$ and in this model the Hubble rate $\tilde{H}$ diverges at $\tilde{t}=0$, so we have a Type III singularity in the Einstein frame.

By using the standard procedure outlined above, we can apply a conformal transformation to convert the theory to the Jordan frame. The new time coordinate $t$ can be found by solving the following differential equation,
\begin{align}
\frac{dt}{d\tilde{t}}=(\tilde{t}/\tilde{t}_0)^{\pm\sqrt{\frac{p}{3}}}\,,
\end{align}
which has the following solution,
\begin{align}
t= \frac{3 }{3\pm \sqrt{3p}} \, \tilde{t} \left(\frac{\tilde{t}}{\tilde{t}_{0}}\right)^{\pm2\sqrt{\frac{p}{3}}}\,.
\end{align}
The corresponding scale factor as a function of the cosmic time reads,
\begin{align}\label{scalej}
a(t)\sim  t^{\frac{\sqrt{3p}\pm3 p}{\sqrt{3p}\pm 3}}\, .
\end{align}
In the case where the minus sign is chosen in the conformal factor, the cosmological evolution~(\ref{scalej}) has a Type I finite time singularity at $t=0$ if the power law parameter $p$ lies in the range $1/3\leq p<3$. If $p=1/3$, the Jordan frame metric becomes static and there is no longer a singularity. In all other cases, the Type III singularity at $\tilde{t}=0$ in the original Einstein frame remains a type III singularity at $t=0$ in the Jordan frame.

\subsection{Cosmology Generated by $R^{-n}$ Gravity in the Jordan Frame}

We consider now the cosmology corresponding to an $R^{-n}$ gravity in the Jordan frame, which was studied in detail in ~\cite{Briscese:2006xu}. When $F(R)$ behaves as $F(R)\sim R^{-n}$, it can be seen from~(\ref{JGRG15}) and~(\ref{Cr4b}) that the corresponding scale factor behaves as the following power law type,
\begin{align}
a\sim \left(t_0 - t\right)^{\frac{(n+1)(2n+1)}{n+2}}\ .
\end{align}
Therefore, if either $n<-2$ or $-1<n<-1/2$, a Big Rip Type I singularity appears at the time instance $t=t_0$ in the Jordan frame, and in the remaining cases a Type III Big Crunch singularity is present at this point. In this case, the corresponding Einstein frame canonical scalar field reads,
\begin{align}
\sigma\sim (n+1)\ln R \sim -2(n+1)\ln (t_0 - t)\ ,
\end{align}
where we took into account that the Ricci scalar reads,
\begin{align}
R\sim \frac{6(n+1)(2n+1)(4n+5)n}{(n+2)^2(t_0 - t)^2}\ .
\end{align}
In the corresponding scalar-tensor theory, the time coordinate $\tilde{t}$ is given by, 
\begin{equation}\label{sctefr}
d\tilde t=\pm e^{\frac{1}{2}\sigma}dt\sim \pm (t_0 - t)^{-(n+1)}dt\, ,
\end{equation}
and consequently, we have $\tilde t=\pm (t_0-t)^{-n}$. Therefore, in the case that $n>0$, when $t$ approaches $t\to t_0$ in the Jordan frame, this corresponds to $\tilde{t}\to \pm \infty$ in the Einstein frame. As a consequence, the singularity changes its structure, since it does not appear in finite time for the scalar-tensor theory, however a new additional singularity may be present, as when $t$ approaches infinity in the Einstein frame, it corresponds to the new time coordinate $\tilde{t}\to 0$, and thus any singularities at infinity can be brought towards a finite time. On the other hand, when $n<0$, the limit $t\to t_0$ in the Jordan frame corresponds to $\tilde{t}\to 0$ in the Einstein frame. We also find that the metric in the scalar-tensor theory behaves as
\begin{align}
ds_{ST}^2=e^{\sigma} \left(-dt^2 + a(t)^2\sum_{i=1,2,3}(dx^i)^2\right)
\sim -d\tilde t^2 + \tilde a(\tilde{t})^2 \sum_{i=1,2,3}(dx^i)^2\ ,\quad
\tilde a(\tilde{t})^2 \sim a_0^2 \tilde t^{\frac{2n(n^2-1)}{n+2}}\ ,
\end{align}
where the constant $a_0$ is an arbitrary parameter. In this case the power of the scale factor is negative only when $-2<n<-1$ or $0<n<1$, and thus a Big Rip Type I singularity becomes present then. Thus for the Big Rip Type I singularity in the Jordan frame, the scale factor now behaves as $\tilde{a}(\tilde{t})^2\to 0$ when $\tilde t\to 0$, in the Einstein frame it becomes a Type III Big Crunch singularity. 

\subsection{A Singular Cosmological Evolution}

In this section we investigate how the simplest singular cosmology, behaves in the corresponding frame. The simplest singular cosmology is described by the following Hubble rate,
\begin{align}\label{singein}
H(\tilde{t})=f_0(\tilde{t}-\tilde {t_s})^\alpha\,,
\end{align}
with $f_0$ an arbitrary real and positive parameter and $\alpha$ a real number, the values of which will determine the singularity Type. Particularly, depending on the values of the parameter $\alpha$, we have the following singularities in the cosmological evolution,
\begin{itemize}
\item When $\alpha<-1$, the cosmology develops a Type I singularity.
\item When $-1<\alpha<0$, the cosmology develops a Type III singularity.
\item When $0<\alpha<1$, the cosmology develops a Type II singularity.
\item When $\alpha>1$, the cosmology develops a Type IV singularity.
\end{itemize}
Assume that the cosmological evolution (\ref{singein}), occurs in the Einstein frame, and the question is what happens if we convert to the Jordan frame $F(R)$ theory? The conformal factor is given by $e^{\sqrt{\frac{2}{3}}\phi}$ and therefore the metric transforms as follows,
\begin{align}
ds_{F(R)}^2=e^{\sqrt{\frac{2}{3}}\phi}\left(-d\tilde{t}^2+ \tilde{a}(\tilde{t})^2\sum_{i=1,2,3} \left(dx^i\right)^2 \right) \, ,
\end{align}
consequently, the scale factor becomes,
\begin{align}
a(t)=e^{\frac{1}{2}\sqrt{\frac{2}{3}}\phi}\tilde{a}(\tilde{t})\, ,
\end{align}
where the Jordan frame time parameter $t$ is defined implicitly through the following relation,
\begin{align}
dt=e^{\frac{1}{2}\sqrt{\frac{2}{3}}\phi} d\tilde{t}\, .
\end{align}
The solution to this differential equation is an incomplete gamma function. Now if the Hubble rate $H$ is described by Eq.~(\ref{singein}), then from the corresponding equations of motion, the Einstein frame scalar field is equal to,
\begin{align}
\phi=\frac{2 \sqrt{-2f_{0}\alpha } (\tilde{t}-\tilde{t}_s)^{\frac{\alpha +1}{2}}}{\alpha +1}\,,
\end{align}
where $f_0 \alpha<0$ is required in order for the scalar field to be canonical. The transformation blows up at $\tilde{t}_s$ if $\alpha<-1$, so extra caution is required in this case. In effect, the new Hubble rate in terms of our initial time coordinate $\tilde{t}$ is equal to,
\begin{align}
H(\tilde{t})=\frac{ \sqrt{-f_0\alpha} (\tilde{t}-\tilde{t}_s)^{\frac{\alpha -1}{2}}}{\sqrt{3}}+f_0 (\tilde{t}-\tilde{t}_s)^{\alpha }\, ,
\end{align}
and the question is what is the effect on the time coordinate $t$. Suppose that, $t=f(\tilde{t})$. When, $f(\tilde{t})\neq 0$, we have,
\begin{align}
f'(\tilde{t})=e^{\frac{1}{2}\sqrt{\frac{2}{3}}\phi}\, ,
\end{align}
and correspondingly, the following relations holds true,
\begin{align}
\frac{dH}{dt}=\frac{d\tilde{t}}{dt} \frac{d\tilde{H}}{d\tilde{t}}=e^{-\frac{1}{2}\sqrt{\frac{2}{3}}\phi}\frac{d\tilde{H}}{d\tilde{t}}\, .
\end{align}
This means that the expression $\frac{dH}{dt}$ diverges if and only if $\frac{d\tilde{H}}{d\tilde{t}}$ diverges for $\alpha>-1$, as then the conformal factor is finite at $\tilde{t}_s$. Effectively, a Type I singularity occurs when the Hubble rate $H(t)$ diverges and the question is whether the singularity still appears at a finite time, we need to investigate whether $f(\tilde{t}_s)$ is finite. It is easy to show that,
\begin{align}
t_s=f(\tilde{t}_s)=c_1-c_2 \Gamma \left(\frac{2}{\alpha +1}\right)\, ,
\end{align}
which is finite, provided $\frac{2}{1+\alpha}$ is not a negative integer. Consequently, the singularity appears in the Hubble rate $H$ at a finite time, as long as $\alpha \neq 2/n-1$ where $n\geq 2$ is an integer. Accordingly, a Type II singularity occurs if $\frac{dH}{dt}$ diverges, but $H$ does not diverge. Combined together, these imply that $1<\alpha<3$. Note that for $\alpha>3$ a Type IV singularity occurs. Now we investigate how the scale factor behaves when it is conformally transformed, which in the Einstein frame reads,
\begin{align}
\tilde{a}(\tilde{t})=Ce^{\frac{f_0(\tilde{t}-\tilde{t}_s)^{1+\alpha}}{1+\alpha}}\, ,
\end{align}
so the conformally transformed scale factor in the Jordan frame reads,
\begin{align}\label{scfjf}
a(\tilde{t})=a_{0} e^{ \left(\frac{3 f_{0} (\tilde{t} -\tilde{t}_{s})^{\alpha+1}\pm 2 \sqrt{-3\alpha f_{0}}  (\tilde{t} -\tilde{t}_{s})^{\frac{\alpha+1}{2}}}{3 (\alpha+1)}\right)}\,,
\end{align}
with $\beta$ an arbitrary constant. The scale factor (\ref{scfjf}) dictates the following singularity patter for the cosmological evolution, depending on the values of the parameter $\alpha$,
\begin{itemize}
    \item For $\alpha<-1$, a Type I or no singularity occurs.
    \item For $-1<\alpha<1$, a Type III singularity occurs.
    \item For $1<\alpha<3$, a Type II singularity occurs.
    \item For $3<\alpha$, a Type IV singularity occurs.
\end{itemize}
In Table \ref{firsttable} we have presented the singularity correspondence for the Jordan and Einstein frame. As we can see, the most interesting cases are the Type I, Type II and Type IV singularities, since the case of the Big Rip singularity in the Einstein frame may correspond to a non-singular evolution in the Jordan frame. Also the Type II singularity can be modified to a more severe Type III singularity in the Jordan frame, and the Type IV in the Einstein frame can correspond to a Type II singularity in the Jordan frame.

\begin{table*}[h]
    \small
\begin{tabular}{@{}|c|r|rrrrrrrrrr@{}}
        \tableline
        \tableline
        \tableline
        Singularity in Einstein Frame & Singularity in Jordan Frame 
        \\\tableline
        Type I & Type I or no singularity 
        \\\tableline
        Type III & Type III 
        \\\tableline
        Type II & Type III
        \\\tableline
        Type IV & Type IV or Type II 
        \\\tableline
        \tableline
    \end{tabular}
    \caption{\label{firsttable}Correspondence for finite time singularities in the Einstein and Jordan frames, for the cosmological evolution $\tilde{H}(\tilde{t})=f_0(\tilde{t}-\tilde{t}_s)^{\alpha}$ in the Einstein frame.}
\end{table*}

A special case of the singular evolution (\ref{singein}) is the singular bounce cosmology which is a special case of the symmetric bounce \cite{unimodular1}. We can rewrite the scale factor of the cosmological evolution (\ref{singein}), in the following form,
\begin{align}\label{singbounce}
a(t)=e^{f_0(t-t_s)^{2(1+\epsilon)}}\, ,
\end{align}
in which case, the Hubble rate reads,
\begin{align}\label{hubrat1}
H(t)=2(1+\epsilon)f_0 (t-t_s)^{2\epsilon+1}
\end{align}
where $\epsilon>0$ and has to be carefully chosen so that everything remains real. Particularly, in order for a bounce to occur, for $t<t_s$ the Hubble rate must become negative, that is $H<0$ and also in order for the bounce (\ref{singbounce}) to be a deformation of the symmetric bounce $a(t)\sim e^{\beta t^2}$, the parameter $\epsilon$ must be chosen in the interval $0<\epsilon<1$ and also must be of the following form,
\begin{equation}\label{epsilon}
\epsilon=\frac{2n}{2m+1}\, ,
\end{equation}
where $m$ and $n$ are integers chosen in such a way so that $\epsilon<1$. Clearly, for this choice of $\epsilon$, the cosmology described by the scale factor (\ref{singbounce}) and the Hubble rate (\ref{hubrat1}) clearly describes a Type IV singular cosmology, in which case, the Hubble rate and its first derivative $\dot{H}$ are finite, but the second derivative with respect to the cosmic time $\ddot{H}$ diverges. As was demonstrated in \cite{noo3}, by using well known reconstruction techniques \cite{reviews1}, the pure $F(R)$ gravity which can realize the cosmological evolution (\ref{hubrat1}) is approximately given by,
\begin{align}\label{frgr}
F(R)= R+ \frac{R^2}{4C_0}+\Lambda\,,
\end{align}
where $C_0$ is positive, near the bouncing point, which is $t\simeq t_s$. For simplicity we introduce the parameter $x=t-t_{s}$, so the limit near the bouncing point corresponds to the limit $x\rightarrow 0$. In order to transform the theory in the Einstein frame, we perform the following conformal transformation,
\begin{align}
g_{\mu\nu}=e^{-\sigma} \hat{g}_{\mu\nu}\, ,
\end{align}
where the scalar field $\sigma$ is equal to,
\begin{align}
\sigma= \ln F'(A)\, .
\end{align}
In terms of the parameter $x$, the Ricci scalar is given by,
\begin{align}
R=12 f_0 (\epsilon +1) x^{2 \epsilon } \left(4 f_0 (\epsilon +1) x^{2 \epsilon +2}+2 \epsilon +1\right)
\end{align}
and so if we are close to the singularity, in which case $x$ is small, the Ricci scalar becomes,
\begin{align}\label{riccisc}
R\approx 12 f_0 (\epsilon +1)(2\epsilon+1) x^{2 \epsilon } \, .
\end{align}
Consequently, by combining Eqs. (\ref{frgr}) and (\ref{riccisc}), we obtain,
\begin{align}
F'(R)\approx 1+\frac{6 f_0 (\epsilon +1)(2\epsilon+1) x^{2 \epsilon } }{C_0}=1+C_{2} x^{2\epsilon}\, ,
\end{align}
where $C_{2}$ is another positive constant. This means that the new time coordinate of the Einstein frame FRW metric will be given in terms of $x$ as,
\begin{align}\label{diffeq1}
\mathrm{d}\tilde{t}=(1+C_{2} x^{2\epsilon})^{\frac{1}{2}\sqrt{\frac{3}{2}}}\mathrm{d}x\, ,
\end{align}
the solution to which is a hypergeometric function. The new scale factor in terms of $x$ is given by,
\begin{align}
a(\tilde{t})=(1+C_{2} x^{2\epsilon})^{\frac{1}{2}\sqrt{\frac{3}{2}}} \tilde{a}(x)\, ,
\end{align}
and consequently, the derivative of the scale factor is given by
\begin{align}
\frac{d a}{d\tilde{t}}= \frac{dx}{d \tilde{t}} \frac{d a}{dx}= \frac{dx}{d\tilde{t}}\left((1+C_{2} x^{2\epsilon})^{\frac{1}{2}\sqrt{\frac{3}{2}}} \frac{d\tilde{a}}{dx}+2\epsilon x^{2\epsilon}(1+C_{2} x^{2\epsilon})^{\frac{1}{2}\sqrt{\frac{3}{2}}-1} \tilde{a}(x)\right)\, .
\end{align}
Owing to the fact that, $\mathrm{d}t=\mathrm{d}x$ at $x\simeq 0$, by looking at the second derivative of the scale factor $a(t)$ it easily follows that it diverges at $x=0$ provided $\epsilon<1/2$. Thus the Type IV singularity in the Jordan frame becomes a Type II singularity in the Einstein frame, which is expected according to Table \ref{firsttable}.

\section{Unimodular $F(R)$ gravity}\label{unimodularfr}

Apart from the standard $F(R)$ gravity approach, which we developed in the previous sections, we shall study the correspondence between frames in the context of unimodular $F(R)$ gravity, with the latter being developed in Refs. \cite{unimodular1,unimodular4}. The general action of unimodular $F(R)$ gravity reads \cite{unimodular1},
\begin{align}
    \label{Uni11}
    S = \int d^4 x \left\{\sqrt{-g} \left( F(R) - \lambda \right) + \lambda \right\}
    + S_\mathrm{matter} \,,
\end{align}
with $F(R)$ being a smooth function of the Ricci scalar, $\lambda$ is the Lagrange multiplier function and $S_\mathrm{matter}$ stands for the action of the matter fluids present. If we do a variation with respect to $\lambda$, we obtain the unimodular constraint, 
\begin{align}
    \sqrt{-g}&=1\,,\label{condition}
\end{align}
so that the determinant of the metric is fixed. The unimodular constraint is the central point of unimodular $F(R)$ gravity. If we vary the action (\ref{Uni11}) with respect to the metric tensor $g_{\mu \nu}$, we obtain the unimodular $F(R)$ field equations given by,
\begin{align}
    \frac{1}{2}g_{\mu\nu} \left( F(R) - \lambda \right) - R_{\mu\nu} F'(R) 
    + \nabla_\mu \nabla_\nu F'(R) - g_{\mu\nu}\nabla^2 F'(R) + \frac{1}{2} T_{\mu\nu}&=0. \, \label{F(R)2}
\end{align}
 In order to study cosmology, one needs to be very careful since the flat standard FRW metric of Eq. (\ref{metricfrw})
does not satisfy the unimodular constraint (\ref{condition}). However, upon making the following coordinate transformation,
\begin{align}
    d\tau&=a(t)^3dt,\label{dtaudt}
    \end{align}
the resulting metric satisfies the unimodular condition (\ref{condition}). Using the transformation (\ref{dtaudt}), we obtain,
\begin{align}
    \label{UniFRW2}
    ds^2 = a\left(t\left(\tau\right)\right)^{-6} d\tau^2 - a\left(t\left(\tau\right)\right)^{2} \Big(dx^2+dy^2+dz^2\Big) \, .
\end{align}
which we will call for brevity, the unimodular FRW metric. Using this metric, the vacuum field equations become \cite{unimodular1},
\begin{align}
    \label{Uni14}
    0 = & - \frac{a^{-6}}{2} \left( F(R) - \lambda \right) + \left( 3 \dot K + 12 K^2 \right) F'(R) 
    - 3 K \frac{d F'(R)}{d\tau}  \, , \\
    \label{Uni15}
    0 = & \frac{a^{-6}}{2} \left( F(R) - \lambda \right) - \left( \dot K + 6 K^2 \right) F'(R) 
    + 5 K \frac{d F'(R)}{d\tau} + \frac{d^2 F' (R)}{d\tau^2}  \, ,
\end{align}
where the function $K(\tau)$ is defined to be the corresponding Hubble rate in the ``$\tau$'' coordinate, that is,
\begin{align}
K=\frac{1}{a(\tau)}\frac{d a(\tau)}{d\tau}.
\end{align}
By using the unimodular FRW metric of Eq. (\ref{UniFRW2}), the corresponding Ricci scalar reads,
\begin{align}
R&=a^{6}(6\dot{K}+30K^2).
\end{align}
In the following sections we shall study the correspondence of the Jordan frame unimodular $F(R)$ gravity in the Einstein frame.

\subsection{Scalar-tensor Einstein Frame Representation}
\label{sec:Unimodularconformal}

Having presented the unimodular $F(R)$ gravity in the Jordan frame, we now study the corresponding scalar-tensor theory. We start off with the action of Eq. (\ref{Uni11}), omitting the matter fluids, and we introduce the auxiliary field $A$, as we did in the ordinary $F(R)$ gravity case, so that the action becomes,
\begin{align}\label{atscaction}
S = \int d^4 x \left( \sqrt{-g} \left( F'(A)(R-A)+F(A) - \lambda \right) + \lambda \right)\, .
\end{align}
Note that the last term of the action will be unaffected by conformal transformations. In order to obtain a minimally coupled scalar-tensor theory, we perform the same conformal transformation we performed in the standard $F(R)$ gravity case, which is,
\begin{align}
\hat{g}_{\mu\nu}= e^\sigma g_{\mu\nu}\,,
\end{align}
where $\hat{g}$ denotes the metric in the Einstein frame, and $g$ the metric in the Jordan frame. Also, the scalar field $\sigma$ in terms of $A$ is given by, $\sigma=-\ln F'(A)$, so the action (\ref{atscaction}) becomes,
\begin{align}
S = \int d^4 x \left\lbrace \sqrt{-\hat{g}} \left( \hat{R}-\frac{3}{2}\hat{g}^{\mu\nu}\partial_{\mu}\sigma \partial_\nu \sigma-V(\sigma) - \lambda e^{-2\sigma}\right) + \lambda  \right\rbrace\, ,
\end{align}
which describes a canonical scalar field action, in the absence of any matter fluids. Note however that the unimodular constraint is not unaffected by the conformal transformation, to in the case at hand it becomes, 
\begin{align}\label{confrelated}
\sqrt{-\hat{g}}=e^{2\sigma}\, .
\end{align}
In effect, the FRW metric of Eq. (\ref{metricfrw}) does not satisfy this constraint identically, so in the same way we evades this issue previously, we introduce a new time coordinate $\tau$, which is related to the cosmic time $t$ as follows,
\begin{align}
d\tilde{\tau}=\tilde{a}(\tilde{t})^3 e^{2\sigma(\tilde{t})}d\tilde{t}\, ,
\end{align}
so that the conformally transformed unimodular constraint of Eq. (\ref{confrelated}) is satisfied. The corresponding Einstein frame unimodular FRW metric becomes, 
\begin{align}
\label{conformalmetricscalar}
ds^2=-\frac{e^{4\sigma(\tilde{\tau})}}{\tilde{a}(\tilde{\tau})^6}d\tilde{\tau}^2+\tilde{a}(\tilde{\tau})^2\sum_{i=1}^3dx_i^2 \, .
\end{align}

To transform between the two frames at the level of the metric then, the following transformation happens. The scale factor transforms as
\begin{align}\label{scaletransform}
\tilde{a}(\tilde{\tau})&=e^{\sigma/2} a(\tau)\,,
\end{align}
where the parameter $\tilde{\tau}$ is related to $\tau $ as follows,
\begin{align}\label{scaletransform2}
\frac{e^{4\sigma(\tilde{\tau})}}{\tilde{a}(\tilde{\tau})^6}d\tilde{\tau}^2&=\frac{e^\sigma}{a(\tau)^6}d\tau^2\,.
\end{align}
By combining Eqs. (\ref{scaletransform}) and (\ref{scaletransform2}), we can easily see that the new time coordinate $\tilde{\tau}$ is the same as the coordinate $\tau$, that is, $\tilde{\tau}=\tau$.

\subsection{Jordan frame of Canonical Scalar Field Model}
Up to now we have discussed the unimodular $F(R)$ gravity case, but we also need to discuss what is the scalar-tensor unimodular gravity. Let us start with the following minimally coupled scalar-tensor action,
\begin{align}
S = \int d^4 x \left\lbrace \sqrt{-\hat{g}} \left( \frac{\hat{R}}{2\kappa^2}-\frac{1}{2}\hat{g}^{\mu\nu}\partial_{\mu}\phi \partial_\nu \phi-V(\phi) - \lambda h(\phi)\right) + \lambda  \right\rbrace\,,
\end{align}
where for now we have assumed that the determinant of the metric is given by an arbitrary function of the scalar field. This will be determined later by the requirement that the action has a Jordan frame. The flat unimodular FRW equations for this action are given by the standard scalar field cosmological equations with a modified scalar potential of the form $V(\phi) +\lambda h(\phi)$. Explicitly, we have
\begin{align}
3\tilde{K}^{2}&=\frac{1}{2}\dot{\phi}^{2}+\Big(V(\phi) +\lambda h(\phi)\Big)\tilde{a}(\tilde{\tau})^{-6}\,,\label{EQ1}\\
9\tilde{K}^2+2\dot{\tilde{K}}&=-\frac{1}{2}\dot{\phi}^{2}+\Big(V(\phi) +\lambda h(\phi)\Big)\tilde{a}(\tilde{\tau})^{-6}\,.\label{EQ2}
\end{align}
Here $\tilde K(\tilde\tau)$ represents the unimodular Hubble parameter in the Einstein frame, explicitly given by
\begin{align}
\tilde K(\tilde\tau)&=\tilde K(\tau)=\frac{1}{\tilde{a}(\tau)}\frac{d\tilde a(\tau)}{d\tau}\,.
\end{align}

Now let us attempt to conformally transform this action to a Jordan frame. We apply the following conformal transformation,
\begin{align}
g_{\mu\nu}=e^{\pm\kappa\sqrt{\frac{2}{3}}\phi}\hat{g}_{\mu\nu}\label{transfor}\, .
\end{align} 
This rescaling eliminates the kinetic term from the action, reducing it to the following form,
\begin{align}
S = \int d^4 x \left\lbrace \sqrt{-g} \left( e^{\pm\kappa\sqrt{\frac{2}{3}}\phi}{2\kappa^2}R-e^{\pm2\kappa\sqrt{\frac{2}{3}}\phi}(V(\phi) + \lambda h(\phi))\right) + \lambda  \right\rbrace\,.
\end{align}
Varying this action with respect to the scalar field $\phi$, which is now just an auxiliary field, gives the following equations of motion,
\begin{align}
R=e^{\pm\kappa\sqrt{\frac{2}{3}}\phi}\left( 4\kappa^2 (V(\phi)+\lambda h(\phi))\pm2\kappa \sqrt{\frac{2}{3}}(V'(\phi)+\lambda h'(\phi))\right)\,.
\end{align}
To find the Jordan frame unimodular $F(R)$ gravity, we need to invert this equation to find the scalar field $\phi (R)$, as a function of $R$ only, and so independent of $\lambda$. This determines the up to now arbitrary function $h(\phi)$, which must be chosen as follows,
\begin{align}
h(\phi)=e^{-2\kappa\sqrt{\frac{3}{2}}\phi}\, .
\end{align}
In this case we can invert to find $\phi=\phi(R)$ as usual, so the resulting unimodular $F(R)$ gravity theory reads,
\begin{align}
S = \int d^4 x \left\lbrace \sqrt{-g} \left( F(R) + \lambda \right) + \lambda  \right\rbrace\,,
\end{align}
where the $F(R)$ gravity is,
\begin{align}
F(R)= e^{\pm\kappa\sqrt{\frac{2}{3}}\phi(R)}{2\kappa^2}R-e^{\pm 2\kappa\sqrt{\frac{2}{3}}\phi(R)}V(\phi(R)).
\end{align}

Now that we have the corresponding Jordan and Einstein frames, in the following sections repeat the analysis performed in the standard $F(R)$ case and present some illustrative examples to see the correspondence of finite time singularities in the two frames. Again, all quantities in the Einstein frame will be denoted with a tilde, and we will set the gravitational coupling constant $\kappa=1$.

\subsection{Power law}
Let us begin by examining how a power law scale factor transforms when one conformally maps from the Einstein frame to the Jordan frame. Let us consider the following scale factor in terms of the time coordinate $\tilde{\tau}$ from the unimodular Einstein frame FRW metric~(\ref{conformalmetricscalar}) 
\begin{align}
\tilde{a}(\tilde{\tau})&=a_{0}\Big(\frac{\tilde{\tau}}{\tilde{\tau}_{0}}\Big)^{p}\, .
\end{align}
Here $\tilde{\tau}_{0}$ is some fiducial time and $p$ is a positive parameter. Using this scale factor, the unimodular Hubble parameter is $\tilde K(\tau)=p\tilde{\tau}^{-1}$. Therefore at $\tilde{\tau}=0$, the Hubble rate diverges and hence this cosmology possesses a Type III singularity in the Einstein frame. Such a scale factor is a solution to the Einstein frame Friedmann equations~(\ref{EQ1})-(\ref{EQ2}) when the potential is of the exponential type. In this case the scalar field takes the form 
\begin{align}
\phi(\tilde{\tau})=\pm\sqrt{2p(1-3p)}\,\log(\tilde{\tau}/\tilde{\tau}_{0})\,.\label{scalefactor2}
\end{align} 

Using this solution, we can now change all our variables from the Einstein frame to the Jordan frame, using the conformal transformation
\begin{align}
g_{\mu\nu}=e^{\pm\kappa\sqrt{\frac{2}{3}}\phi}\hat{g}_{\mu\nu}\label{transform2}\,. 
\end{align}
As we discussed in Sec.~\ref{sec:Unimodularconformal}, the time coordinate $\tau$ in the Jordan frame is equivalent to the original time coordinate in the Einstein frame.  So using the transformation~(\ref{transform2}), we find that the scale factor in the Jordan frame is given by
\begin{align}
a(\tau)=e^{\pm\frac{1}{2}\sqrt{\frac{2}{3}}\phi}\tilde{a}(\tilde{\tau})\sim  \tau ^{p\pm\sqrt{\frac{p(1-3 p)}{3}}} \,.\label{scalefactorpower}
\end{align}
When the minus sign is chosen for the conformal transformation, we see that there is a potential for the Type III singularity to become a Big Rip Type I singularity. This can happen at $\tau=0$ if the power law parameter lies in the range $0<p< 1/6$. For all other cases the Type III singularity remains a Type III singularity in the Jordan frame.

\subsection{The case $F(R)=R^{-n}$ Singularity Types in the Jordan Frame}

Now let us examine a model where we begin in the Jordan frame and conformally transform to the Einstein frame. We consider vacuum unimodular $F(R)$ gravity of the form $F(R)\sim R^{-n}$, in which case the scale factor behaves as follows,
\begin{align}
a(\tau)\sim(\tau_0-\tau)^{\frac{1+3n+2n^2}{5+10n+6n^2}}\,,
\end{align}
where we used the $\tau$ coordinate of the unimodular FRW metric. It is obvious that a Big Rip singularity occurs in the $\tau$ coordinate, if the parameter $n$ lies in the range $-1<n<-1/2$. In terms of the original FRW cosmic time coordinate $t$, this means the scale factor evolves as,
\begin{align}
a(t)\sim t^{\frac{(2n+1)(n+1)}{2+n}}\,.
\end{align}
Thus a Big Rip singularity in the $\tau$ coordinate, appears in the $t$ coordinate if $n<-2$ or $-1<n<-1/2$.

The question now is what happens in the corresponding Einstein frame scalar-tensor theory. The scalar field $\sigma$ of the conformal factor is given by
\begin{align}
\sigma \sim (n+1)\ln R \sim -\left(\frac{2(n+1)(2+n)}{5+10n+6n^2}\right)\ln (\tau_0-\tau),
\end{align}
where we use the fact that the Ricci scalar in the unimodular time parameter is given by
\begin{align}
R\sim (\tau_0-\tau)^{-2\left(\frac{2+n}{5+10n+6n^2}\right)}.
\end{align}
This means under a conformal transformation, the scale factor transforms as follows
\begin{align}\label{scale1}
\tilde{a}(\tilde{\tau})&=e^{\sigma/2} a(\tau)
\end{align}
where the parameter $\tilde{\tau}$ is the same as the original time coordinate $\tau $, $\tilde{\tau}=\tau$. Therefore we have,
\begin{align}
    \tilde{a}\sim (\tau_{0}-\tau )^{\frac{n^2-1}{6 n^2+10n+5}}\,.
\end{align}
These conditions enlarge the range of values of $n$ for which the power of the scale factor is negative, with now $n$ lying in the range $-1<n<1$ giving rise to this. Thus a Type I Big Rip will appear for more values of $n$.

\subsection{A singular cosmological model}
As a final example, we will consider a toy model where the unimodular Hubble parameter in the Einstein frame is
\begin{align}
\tilde K(\tilde{\tau})&=f_{0}(\tilde{\tau}-\tilde{\tau}_{s})^{\alpha}\,, \label{toyunimodular}
\end{align}
where $f_{0}$, $\alpha$ are real constants and $\tilde{\tau}_{s}$ is some time. In this case, the corresponding scale factor is
\begin{align}
\tilde a(\tilde{\tau})&=a_{0} e^{\frac{f_{0} (\tilde{\tau}-\tilde{\tau}_{s})^{\alpha+1}}{\alpha+1}}\,,
\end{align}
where $a_{0}$ is a constant.

Now, in order to convert this Einstein frame solution to the Jordan frame, we need to find the scalar field $\phi$ giving rise to such a solution~(\ref{toyunimodular}). To do this we must solve the differential equation obtained by subtracting the unimodular Friedmann equations~(\ref{EQ1}) and ~(\ref{EQ2}). We find
\begin{align}
\dot{\phi}^2+6 f_{0}^2 (\tilde{\tau} -\tilde{\tau}_{s})^{2 \alpha}+2 f_{0} \alpha (\tilde{\tau}-\tilde{\tau}_{s})^{\alpha-1}&=0\,.\label{problem}
\end{align}
This equation can not be integrated in general for an arbitrary power of $\alpha$, and so in order to proceed we will approximate the solution around the singularity at $\tilde{\tau}=\tilde{\tau}_s$. Doing this results in two separate cases: when $\alpha<-1$ and the Hubble rate is that of a Type I singularity, and when $\alpha>-1$ and the other singularity types are present.

In the case when $\alpha<-1$ and then around the singularity, the second term of~(\ref{problem}) dominates over the third term and so, close to the singularity, we can approximate~(\ref{problem}) to be
\begin{align}
\dot{\phi}^2+6 f_{0}^2 (\tilde{\tau} -\tilde{\tau}_{s})^{2 \alpha}&\sim0\,.\label{problem1}
\end{align}
In this case, solving for $\phi$, we find that it becomes imaginary, and thus such a Hubble rate could only be described by a phantom scalar field. For such a phantom field, the corresponding Jordan frame $F(R)$ becomes complex, and so we will not examine this case further.

However, now when $\alpha>-1$, the third term in (\ref{problem}) will dominate over the second term and therefore in this case the scalar field near the cosmological singularity at $\tilde{\tau}=\tilde{\tau}_{s}$ will behave as
\begin{align}
\phi(\tilde{\tau})&\sim \pm \frac{2  \sqrt{-2f_{0}\alpha} }{\alpha+1} (\tilde{\tau} -\tilde{\tau}_{s})^{\frac{\alpha+1}{2}}\, ,
\end{align}
which is real if $-2f_{0}\alpha>0$. And so in this case we can proceed further and conformally transform to the Jordan frame. Applying the conformal transformation~(\ref{transfor}) and using that the time coordinate is unchanged $\tau=\tilde{\tau}$, we find the scale factor in the Jordan frame reads
\begin{align}
a(\tau)&\sim a_{0} e^{ \left(\frac{3 f_{0} (\tau -\tau_{s})^{\alpha+1}\pm 2 \sqrt{-3\alpha f_{0}}  (\tau -\tau_{s})^{\frac{\alpha+1}{2}}}{3 (\alpha+1)}\right)}\,.
\end{align} 

Now from this scale factor we can read off conditions for the different singularity types to exist. We have the following structure
\begin{itemize}
    \item For $-1<\alpha<1$, a Type III singularity occurs.
    \item For $1<\alpha<3$, a Type II singularity occurs.
    \item For $3<\alpha$, a Type IV singularity occurs.
\end{itemize}
Table~\ref{secondtable} shows a summary of how the different singularities change type from one frame to another. The Type I singularity is excluded from this table, since it only occurs when $\alpha<-1$ and the scalar field becomes a phantom. We observe that the unimodular $F(R)$ case behaves in a very similar way to the standard $F(R)$ case. The most interesting cases are the Type II and the Type IV singularities in the Einstein frame. The Type II singularity is modified to the more severe Type III singularity, with the Hubble rate now diverging. There is also the potential for the Type IV singularity to become a more severe Type II singularity if the parameter $\alpha$ lies in the range $1<\alpha<3$.

\begin{table*}[h]
    \small
    \begin{tabular}{@{}|c|r|rrrrrrrrrr@{}}
        \tableline
        \tableline
        \tableline
        Singularity in Einstein Frame & Singularity in Jordan Frame 
        %   \\\tableline
        %   Type I & Type I or no singularity 
        \\\tableline
        Type III & Type III 
        \\\tableline
        Type II & Type III
        \\\tableline
        Type IV & Type IV or Type II 
        \\\tableline
        \tableline
    \end{tabular}
    \caption{\label{secondtable}Correspondence for finite time singularities in the Einstein and Jordan frames, for the cosmological evolution  $\tilde{K}(\tau)=f_{0}(\tilde{\tau}-\tilde{\tau}_{s})^{\alpha}$ in the Einstein frame and $\alpha>-1$. }    
\end{table*}

\section{Phase Structure of Unimodular $F(R)$ Gravity near Finite-Time Singularities-A Qualitative Analysis}\label{phasestructurese}

In this section we shall discuss in brief the qualitative behavior of the dynamical system corresponding to the vacuum unimodular $F(R)$ gravity near the finite time singularities. The focus will be exactly for cosmic times near the finite time singularities, so we discuss in brief how the dynamical system behaves near the singularities. However, a most thorough analysis of the dynamical system will be given elsewhere. We start off by introducing the following variables,
\begin{align}
x_1=-\frac{1}{K F'(R)}\frac{dF'(R)}{d\tau}, \quad x_2=-\frac{F(R)}{6K^2 F'(R) a^6}, \quad x_3=\frac{R}{6K^2 a^6}, \quad x_4= \frac{\lambda}{6a^6 K^2 F'(R)}\,.
\end{align}
With this choice, the first Friedmann equation reduces to,
\begin{align}
1=x_1+x_2+x_3+x_4\, . \label{friedmannconstraint}
\end{align}
Thus one can choose to analyze the dynamics of just three of these variables, since one can algebraically relate the fourth to the others. From the second Friedmann equation, we find that,
\begin{align}
\frac{1}{F'(R) K^2}\frac{d^2 F'(R)}{d\tau^2}=1+5x_1+3x_2+x_3+3x_4.
\end{align}
Differentiating with respect to a cosmological time $dN= K(\tau)d\tau$ we obtain the following dynamical system,
\begin{align} \label{dynsystem}
x_1'&=-1+x_1^2-x_1 x_3-3x_2-x_3-3x_4 \\ \notag
x_2'&= -m+4x_2+x_1x_2-2x_2x_3+50-16x_3 \\ \notag
x_3'&= m+20x_3-2x_3^2-50 \\ \notag
x_4'&=x_4(x_1-2x_3+4)
\end{align}
where $m=\frac{\ddot{K}}{K^3}$. Note that the fourth of these equations is superfluous due to the relation~(\ref{friedmannconstraint}), and one can use this to replace $x_4$ in the remaining three equations. Note that the dynamical system, due to the existence of the term $m$, is non autonomous.

The critical points of the system have been displayed in Table~\ref{crit1}. The focus in this section is to investigate certain cases, for which the dynamical system is rendered autonomous. Indeed, if $K(\tau)=f_0(\tau-\tau_s)^{\alpha}$, in the Big Rip case, that is for $\alpha<-1$, for times $\tau\to \tau_s$, the parameter $m$ is equal to, $m=\frac{(-1+\alpha ) \alpha  (\tau -\tau_s)^{-2-2 \alpha }}{f_0^2}\sim 0$, and therefore the dynamical system of Eq. (\ref{dynsystem}) is rendered autonomous for times near the Big Rip singularity. Therefore finding the fixed points of the dynamical system near the Big Rip singularity, may provide some insights on the non-autonomous system, however, we observe that none of the four critical points of the dynamical system are real when one sets $m=0$. In fact, for a constant $m$, one requires that $m\gtrsim 17$ for a critical point to exist. This indicates that the dynamics behave quite strange near the Big Rip singularity but this should be studied in all detail in order to be sure, and the results of this study will be reported elsewhere.

\begin{table}[h]
\begin{center}
\begin{tabular}{|c|c|c|c|c|}
  \hline
  Point & $x_1$ & $x_2$ & $x_3$ & $x_4$    \\
  \hline
  \hline
  $P_1$ & $\frac{1}{2 \sqrt{2}}\left(-\sqrt{m}-\sqrt{m+4 \sqrt{2} \sqrt{m}-40}+2 \sqrt{2}\right)$ & $\frac{1}{4} \left(3 \sqrt{2} \sqrt{m}+\sqrt{2} \sqrt{m+4 \sqrt{2} \sqrt{m}-40}-20\right)$ &$5-\frac{\sqrt{m}}{\sqrt{2}}$ & $0$\\
  \hline
  $P_2$ &$\frac{1}{2 \sqrt{2}}\left(\sqrt{m}-\sqrt{m+4 \sqrt{2} \sqrt{m}-40}+2 \sqrt{2}\right)$ & $\frac{1}{4} \left(3 \sqrt{2} \sqrt{m}-\sqrt{2} \sqrt{m+4 \sqrt{2} \sqrt{m}-40}-20\right)$ &  $5-\frac{\sqrt{m}}{\sqrt{2}}$ & $0$ \\
  \hline
  $P_3$ & $\frac{1}{2\sqrt{2}}\left(\sqrt{m}+\sqrt{m-4 \sqrt{2} \sqrt{m}-40}+2 \sqrt{2}\right)$  & $\frac{1}{4} \left(-3 \sqrt{2} \sqrt{m}-\sqrt{2} \sqrt{m-4 \sqrt{2} \sqrt{m}-40}-20\right)$  &  $5+\frac{\sqrt{m}}{\sqrt{2}}$ & $0$ \\
  \hline
  $P_4$ & $\frac{1}{2\sqrt{2}}\left(\sqrt{m}-\sqrt{m-4 \sqrt{2} \sqrt{m}-40}+2 \sqrt{2}\right)$  & $\frac{1}{4} \left(-3 \sqrt{2} \sqrt{m}+\sqrt{2} \sqrt{m-4 \sqrt{2} \sqrt{m}-40}-20\right)$ & $5+\frac{\sqrt{m}}{\sqrt{2}}$ & $0$ \\
  \hline
\end{tabular}
\end{center}
\caption{Critical points of the dynamical system. }
\label{crit1}
\end{table}
It would be interesting to investigate in detail what is the correspondence of the dynamical systems corresponding to the Einstein and Jordan frames, with regards to singular evolution. The most interesting study is to see the behavior of the dynamical systems near the singularities and compare the behavior of the dynamical systems in both frames. We hope to address this issue in detail in a future work.

\section{Concluding Remarks}

In this work we have considered how the nature of the singularities changes when one conformally transforms between the Jordan and Einstein frames of $F(R)$ gravity. First we considered the Jordan and Einstein frames of classical $F(R)$ gravity and explored various illustrative examples of scale factor evolution. We considered a simple power law potential in the Einstein frame, which exhibits a Type III singularity. However once we transform to the Jordan frame $F(R)$ gravity, we could see that for certain values of the power law parameter the Type III singularity can change its behaviour to become a Type I Big Rip singularity. 

Next we reviewed the situation when in the Jordan frame, $F(R)$ takes the functional form $F(R)=R^{-n}$. In this case the scale factor also behaves as a power law. We find that a Type I Big Rip singularity is removed with the singularity being moved out to infinity, however this introduces a further Type III Big Crunch singularity in the system. Likewise the Type III singularity can be replaced by a Big Rip singularity, depending on the value of the parameter $n$. 

A simple singular cosmology was also considered, where the Hubble rate is described by a power law. Such a toy model is of great use for studying cosmological singularities, since all four types of cosmological singularities can be exhibited by simply changing the power law parameter. We began with such a cosmology in the Einstein frame and conformally transformed to the Jordan frame, and it was found that the type of singularity potentially changes. A Type I singularity was either transformed to another Type I singularity, or the singularity was removed by the time coordinate changing such that the singularity now occurred at infinity. The Type III singularity remained a Type III singularity. On the other hand, a Type II singularity was transformed to the more severe Type III singularity, and likewise the Type IV singularity was potentially transformed to the more severe Type II singularity. 

We also examined the equivalence between the Jordan and Einstein frames of unimodular $F(R)$ gravity. We derived the transformation law to take the canonical Jordan frame unimodular $F(R)$ gravity, where the determinant of the metric is constrained to be $\det (g)=1$, to the Einstein frame. It was found that after conformal transformation, the unimodular constraint was modified accordingly, with the determinant of the transformed metric required to be a particular function of the scalar field. This affects the choice of the time coordinate of the FRW metric required in the Einstein frame. On the other hand, if one starts with an arbitrary unimodular canonical scalar field action, one can only transform to a Jordan frame unimodular $F(R)$ gravity only if the unimodular constraint enforces the determinant of the metric to be this particular function of the scalar field. The behaviour of the unimodular Jordan and Einstein frames should be studied in greater detail, we will leave such an analysis for future work.

We then examined some particular scale factor evolutions and looked at how the structure of the singularities in cosmological times was modified when one transformed between the Jordan and Einstein frames. The power law, $F(R)=R^{-n}$ and the singular toy model were all examined in this framework too. A very similar behaviour to standard $F(R)$ gravity was found, with the singularities transforming to different types of singularities in much the same way. These results could also be readily generalized the framework of other modified $F(R)$ gravities, for example mimetic $F(R)$ gravity~\cite{Nojiri:2014zqa}.  Finally, we briefly examined the cosmological dynamical system of the vacuum unimodular $F(R)$ gravity. We found that close to the Big Rip singularity, the dynamical system exhibits some strange behaviour, with no real critical points present. A further in depth investigation of this phase structure has been left for future work.

\section*{Acknowledgments}
This work is supported by MINECO (Spain), project
 FIS2013-44881 (S.D.O) and by Min. of Education and Science of Russia (S.D.O
and V.K.O). S.B. is supported by the Comisi{\'o}n Nacional de Investigaci{\'o}n
Cient{\'{\i}}fica y Tecnol{\'o}gica (Becas Chile Grant
No.~72150066).

\end{document}